# A Computationally Efficient Model for Pedestrian Motion Prediction

Ivo Batkovic[1,2], Mario Zanon[1], Nils Lubbe[3], and Paolo Falcone[1]

*Abstract*— We present a mathematical model to predict pedestrian motion over a finite horizon, intended for use in collision avoidance algorithms for autonomous driving. The model is based on a road map structure, and assumes a rational pedestrian behavior. We compare our model with the state-of-the art and discuss its accuracy, and limitations, both in simulations and in comparison to real data.

## I. INTRODUCTION

In recent years, autonomous driving has received increased research attention [1], [2], where the main expected benefits include increased safety and efficiency. The first autonomous driving applications will be in structured and easy-to-predict environments such as highway driving and low-speed parking [3], [4]. Arguably, urban scenarios are among the most challenging ones, due to the presence of a large amount of human road users. However, only little research attention has been devoted to deriving models able to predict the stochastic behavior of non-cooperative agents such as human-driven cars, cyclists, and pedestrians.

This paper targets the modeling of pedestrian motion in structured environments, with the aim of using the predicted future pedestrian position in collision-avoidance algorithms. We assume that safety-enforcing strategies manage non road code abiding situations. Therefore, the model is required to accurately predict road code abiding behaviors especially over short horizons. Moreover, since the computations need to be done in the vehicle, a low computational complexity is necessary. Fig.1 shows an example how predicted pedestrian positions can be used for collision avoidance.

Several motion prediction models have been developed based on a variety of different approaches and for different purposes. In [5] a model based on switching between linearizations of a unicycle is used for short prediction horizons up to $2$ s. In [6], hybrid models are used to describe human gaits in order to switch between different dynamics. Situational awareness is introduced on top of switching dynamical models to anticipate changes in pedestrian intent in [7]. Still, this approach has only been applied over short prediction horizons.

We would like to thank Dr. Landgraf and Dr. Erbsmehl from the Fraunhofer Institute for Transportation and Infrastructure Systems IVI, in Dresden, Germany, for the pedestrian dataset. This work was supported by the Wallenberg Autonomous Systems and Software Program (WASP), and by the COPPLAR project (VINNOVA. V.P. Grant No. 2015-04849).
[1] Ivo Batkovic, Mario Zanon and Paolo Falcone are with the Mechatronics group at the Department of Electrical Engineering, Chalmers University of Technology, Gothenburg, Sweden {ivo.batkovic,mario.zanon,falcone}@chalmers.se
[2] Ivo Batkovic is also with the research department at Zenuity AB
[3] Nils Lübbe is with the research department at Autoliv Development AB nils.lubbe@autoliv.com

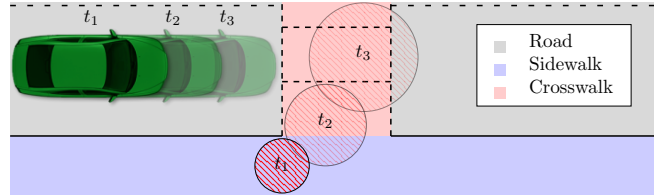

Fig. 1. Example of predicted pedestrian regions (red circles) that the car needs to avoid while driving to ensure safety.

Other approaches where underlying dynamical models were learned using Gaussian processes [8] give accurate predictions for longer time horizons. Predicting motion patterns from data observations also shows improvement in accuracy [9]. However, these approaches have a high computational complexity [10].

Social force models predict behavior of several agents and their interactions. This enables predicting agent movements in crowded environments [11], [12], [13]. The dynamics are often modeled using potential functions to attract or repel the agent from different goals or other agents. However, such methods have been developed for different purposes than autonomous driving and their applicability to traffic scenarios seems limited.

Other methods are based on predefined goals for agents. Modeling agents as moving stochastically along the shortest paths towards these goals leads to accurate long-term predictions [14], [15], [16], [17], [18]. This framework incorporates environmental constraints, e.g regions to avoid or impenetrable objects. In order to obtain the future trajectories, one needs to solve a reinforcement learning (RL) problem, and then sample the learned policy.

This paper aims to provide predictions for collision-avoidance algorithms, therefore we are interested in prediction horizons up to a few seconds. We propose to combine the use of simple strategies, similarly to what has been used in [5], with references related to the road geometry, similarly to the approach used in [17]. Our aim is to benefit from the increased accuracy for longer horizons while maintaining the computational complexity as low as possible. While beyond the scope of this paper, we stress that our approach can be easily combined with situation awareness models, such as those proposed in [7].

This paper is structured as follows. In Section II we introduce the mathematical model. We compare our model to the one proposed in [17] in simulations in Section III and versus real data in Section IV. Over short horizons up to $10$ s we obtain a comparable accuracy for the two models. Finally, we draw conclusions and outline future research directions in Section V.

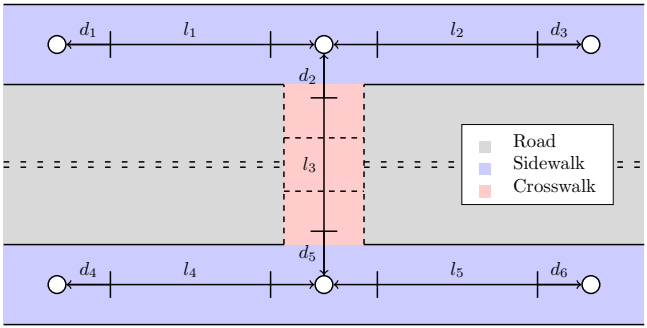

Fig. 2. Example of an intersection: the colored sections refer to different parts of the road; the white circles and the black center lines ($l_i$) depict the nodes and edges of the graph. The black lines around the nodes refer to the distance $d_i$ when the transition between different edges occurs.

## II. PEDESTRIAN MODEL

Pedestrian models to be used for safe autonomous driving need to be computationally efficient but at the same time sufficiently accurate without being overly conservative. In the following, we first introduce the proposed model and subsequently discuss its benefits and limitations.

### A. Mathematical Model

Our model relies on a map of the road configuration, including sidewalks and zebra crossings. All areas where the pedestrian is expected to walk is split into a graph of connected edges, defined so as to yield a connected graph which covers all pedestrian areas of the map. Each edge has an associated reference $\mathbf{r} = [\mathbf{r}^{\mathbf{x}} \ \mathbf{r}^{\mathbf{u}}]^\top$ which is related to the center line of the walking strip. An example of a map is given in Fig. 2.

We model the pedestrian using states $\mathbf{x} \in \mathbb{R}^4$, controls $\mathbf{u} \in \mathbb{R}^2$ and process noise $\mathbf{w} \sim \mathcal{N}(0, W) \in \mathbb{R}^4$, defined as

$$\mathbf{x} := \begin{bmatrix} x \\ y \\ v \\ \theta \end{bmatrix}, \quad \mathbf{u} := \begin{bmatrix} a \\ \omega \end{bmatrix}, \quad \mathbf{w} := \begin{bmatrix} w^x \\ w^y \\ w^v \\ w^\theta \end{bmatrix},$$

where we denote the position coordinates in the absolute reference frame as $x$, $y$ and the velocity in direction $\theta$ as $v$. The acceleration and angular velocity are defined respectively as $a$ and $\omega$. This state representation was used as it explains the physical states at a given time instant. One can alternatively consider lower or higher order models e.g. having the velocity or the jerk as control input.

With the chosen representation, the dynamics are described by the unicycle equations

$$\dot{\mathbf{x}} = \begin{bmatrix} \dot{x} \\ \dot{y} \\ \dot{v} \\ \dot{\theta} \end{bmatrix} = \begin{bmatrix} v\cos\theta \\ v\sin\theta \\ a \\ \omega \end{bmatrix} + \mathbf{w} =: f(\mathbf{x}, \mathbf{u}) + \mathbf{w}. \quad (1)$$

However, for simplicity, we linearize the dynamics around the reference $\mathbf{r}$ of each edge and use a zero-order hold discretization with sampling time $t_s$ to obtain the discrete-time model

$$\mathbf{x}_{k+1} = A\mathbf{x}_k + B\mathbf{u}_k + \mathbf{c}_k + \mathbf{w}_k, \quad (2)$$

where the affine term $\mathbf{c}_k$ accounts for the fact that we linearize the system on a feasible trajectory rather than a steady-state. Note that the model we propose is similar to the one used in [5], with the difference that in our case the system dynamics are linearized around the reference rather than the current state.

In reality, even starting from identical initial conditions, each pedestrian will follow a different trajectory, depending on its own preferences, on the environment, and on the presence of other pedestrians. Such phenomena are impossible to model in detail, and the error introduced by linearization is expected to have a marginal effect on the accuracy of the model. Indeed, our model aims at predicting average behaviors and the covariance of the uncertain trajectories, rather than the behavior of every single pedestrian. In this view, the control $\mathbf{u}$ is used to capture the average behavior, while the process noise $\mathbf{w}$ introduces uncertainty in the model in the attempt to predict the covariance of the future position deviation from the average.

We compute the input as the linear feedback

$$\mathbf{u} = -K(\mathbf{x} - \mathbf{r}^{\mathbf{x}}). \quad (3)$$

Because it is intuitive to tune and understand, we design the feedback matrix $K$ using a linear quadratic regulator (LQR) formulation penalizing the deviation from the reference:

$$\min_{\mathbf{x}, \mathbf{u}} \sum_{k=0}^{\infty} \begin{bmatrix} \mathbf{x}_k - \mathbf{r}_k^{\mathbf{x}} \\ \mathbf{u}_k - \mathbf{r}_k^{\mathbf{u}} \end{bmatrix}^\top \begin{bmatrix} Q & S \\ S^\top & R \end{bmatrix} \begin{bmatrix} \mathbf{x}_k - \mathbf{r}_k^{\mathbf{x}} \\ \mathbf{u}_k - \mathbf{r}_k^{\mathbf{u}} \end{bmatrix} \quad (4a)$$

$$\text{s.t.} \quad \mathbf{x}_0 = \hat{\mathbf{x}}_0, \quad (4b)$$

$$\mathbf{x}_{k+1} = A\mathbf{x}_k + B\mathbf{u}_k. \quad (4c)$$

Using the closed-loop model, we define $A_K = A - BK$ and compute the prediction of average $\bar{\mathbf{x}} := \mathbb{E}[\mathbf{x}]$ and covariance $P := \mathbb{E}[\mathbf{x}\mathbf{x}^\top]$ as

$$\bar{\mathbf{x}}_{k+1} = A_K \mathbf{x}_k, \qquad P_{k+1} = A_K P_k A_K^\top + W. \quad (5)$$

The proposed approach allows us to predict the future behavior while remaining on one edge. In order to transition between different edges, we use the simple rule that, once the pedestrian is sufficiently close to the final node of the edge, the dynamics are propagated along all neighboring edges. In particular, the switch occurs when $\Delta d \leq d$, with

$$\Delta d = (x - \mathbf{r}^x)\cos(\mathbf{r}^\theta) + (y - \mathbf{r}^y)\sin(\mathbf{r}^\theta).$$

Clearly, other criteria could be used. Yet, our approach is simple, and as it will be shown, reasonably accurate.

If more than one edge has a starting node corresponding to the final node of the current edge, i.e. there is a bifurcation, then we propagate the pedestrian predictions along all these edges. This corresponds to assuming that new pedestrians are generated instantaneously at the end of each edge. While this carries no physical meaning, it allows us to make sure that we do account for all possible directions the pedestrian might take.

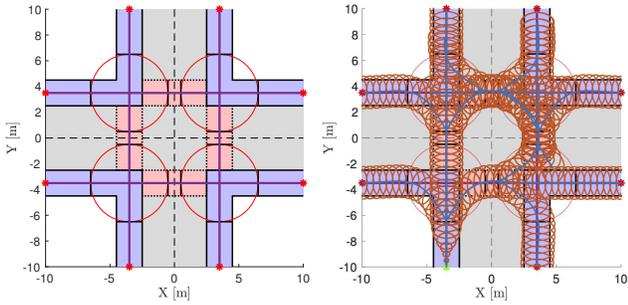

Fig. 3. General intersection. Red points are goals used for the method in [17], whereas the purple lines are the reference paths and the black tangent lines to the circles refer to the minimum distance where switching to new references occurs for our method. The red ellipses correspond to the 99-percentile uncertainty, and the blue lines are the prediction average.

The tuning parameters of the proposed LQR-based model are the stage cost matrices $Q$, $S$, $R$, the process noise covariance $W$ and the distance to the final node $d$. The main advantages of the LQR-based model are the computational simplicity, its intuitive tuning parameters, and its flexibility (since no expensive pre-computation is necessary for different road configurations). The main drawback of the model in its proposed form is that we are able to only predict "nominal", or road code-abiding behavior. However, the model can easily be combined with previous work from [7] to e.g. predict the intent of a pedestrian and add edges to the graph if the possibility of a non-standard behavior is detected. Future research will focus on such extensions of the model.

## III. SIMULATIONS AND QUALITATIVE COMPARISON

In order to motivate the use of the proposed model, we compare our model (5) to the model proposed in [17] in simulations. While other prediction approaches have been developed, we use [17] as a baseline for comparison because, similar to our approach, it uses a model for a stochastic prediction of the pedestrian motion. In contrast to [17], we did not aim at developing long-term motion prediction to study human behavior, but at providing computationally inexpensive short-term predictions.

The model used in [17] is based on RL and uses a unicycle pedestrian model and a semantic map of costs for a stochastic state feedback policy. The presence of bifurcations is handled by introducing goals in the map where the pedestrian is assumed to be willing to go. Given position (and eventually orientation) measurements, a particle filter estimates the probability that the pedestrian will go to each of the available goals. The motion prediction consists in a two-step computation: (a) the RL problem is solved offline once in order to learn the value function for each goal and state, and (b) starting from the current state, the stochastic pedestrian motion is online approximated by sampling based on simulations performed using the policy associated to the learned value function. The tuning parameters of the algorithm are: (a) the cost associated with each region of the map, (b) the discount factor $\gamma$, and (c) parameter $\alpha$ which is related to the inverse of the policy covariance. For more details on the algorithm we refer to [17].

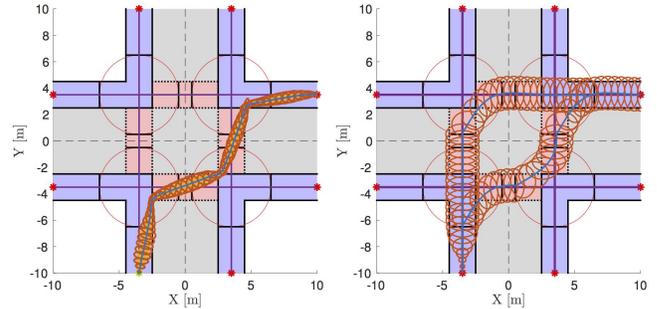

Fig. 4. Predicted motion for the RL-based model [17] (left plot) and the LQR-based model (right plot). Sampled trajectories are displayed in yellow, mean positions are displayed as blue lines, 99-percentile confidence ellipses are displayed in red and the initial state and goal are displayed as green and red asterisks respectively. Note that the covariance ellipses are only drawn for certain time steps in order to improve visibility in the figures.

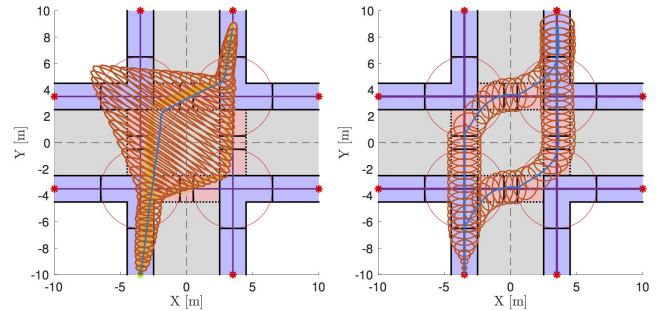

Fig. 5. Scenario yielding conservative naive covariance estimates. The color coding matches the one of Fig. 4.

We consider the simple intersection scenario displayed in Fig. 3, where reference paths (purple lines) connect goals depicted as red asterisks. We generate predictions for the two methods starting from the initial state $\mathbf{x}_0 = [-3.5, -10, 1, \pi/2]^\top$. In the right plot of Fig. 3 we display the 20 s prediction obtained with our model.

In Fig. 4 we focus on the goal $\mathbf{g} = [10, 3.5]^\top$. The results of the RL-based model are displayed in the left plot and those obtained with the proposed model are displayed in the right plot. Note that we display only trajectories for the given goal to simplify visual inspection of the simulation results. The predicted mean state is similar, but the predicted covariance shows a different behavior: for the RL method, its size reduces when approaching corners, while for our method it does not. This difference is based on different use of costs in the feedback policies and reflects the simpler representation of the environment in our approach. We believe that this difference is too small to have practical implications. Furthermore, our estimation is conservative for collision avoidance algorithms, and will not lead to harm.

A major difference between the two approaches occurs when the shortest path towards a given goal is not unique, i.e the road structure is symmetric. Our method handles the bifurcation by branching the prediction in multiple outcomes, whereas the RL method predicts a single multimodal stochastic trajectory. Therefore, a naive covariance computation is bound to yield very conservative estimates, as shown in Fig. 5. While this problem can be solved by either postprocessing the sampling results or by introducing intermediate goals wherever bifurcations can occur, both remedies increase the computational complexity of the algorithm. Lastly, the flexibility of the RL method is limited by the need of computing a value function for every single road configuration. Any changes of the tuning parameters require the same time computations to be redone.

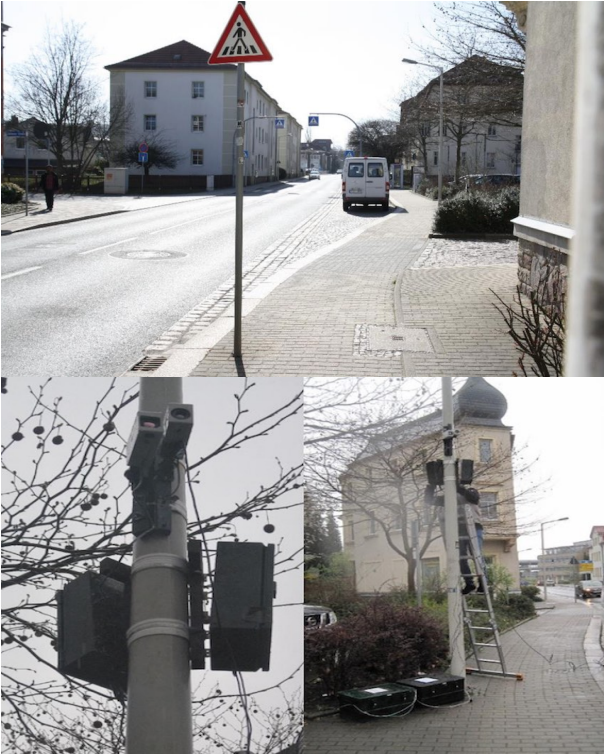

Fig. 6. Intersection and sensor setup.

In this section we compared our model to a state-of-the art model [17]. In the next section, we evaluate the two methods based on data from a pedestrian crossing.

## IV. VALIDATION AGAINST REAL DATA

We use measurements of pedestrians walking near an intersection in Dresden, Germany, displayed in Fig. 6 and 7. The data was captured using an IR camera that was placed in a high position, overlooking the intersection and its sidewalks, while recording at 52 frames per second. The estimation of the pedestrian position was performed by the Fraunhofer Institute for Transportation and Infrastructure Systems IVI in Dresden, Germany, using similar methods and setup as in [19].

The dataset consists of 46 trajectories: 29 crossing the street and 17 remain on the side walks.

### A. Error Metric

We evaluate the accuracy of the model in terms of predicted average and covariance errors. The prediction error for a given trajectory $i$ in the dataset, at time $t$ is

$$\mathcal{E}_{x,i}(t,\tau) = x^{\text{meas}}_{t+\tau} - x_{t+\tau}, \quad (6)$$

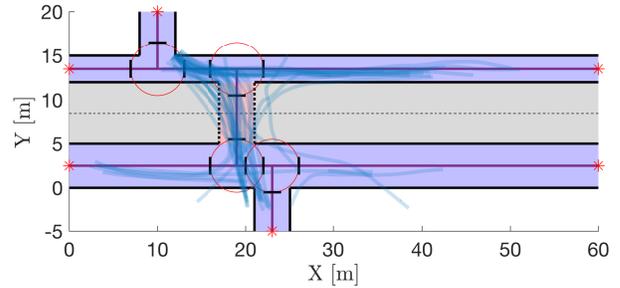

Fig. 7. Intersection map. The notations match the ones of Fig. 3. The transparent blue lines show the measured pedestrian trajectories from the dataset.

where $x_{t+k}$ is the $\tau$ steps-ahead predicted position at time $t$, and $x^{\text{meas}}_{t+k}$ is the corresponding measure. The average error in $x$ is then computed as

$$\hat{\mathcal{E}}_x(\tau) = \frac{1}{N(T-\tau)} \sum_{i=1}^{N} \sum_{t=1}^{T-\tau} \mathcal{E}_{x,i}(t,\tau). \quad (7)$$

The average error $\hat{\mathcal{E}}_y(\tau)$ is defined equivalently. The $\ell_2$ error is then computed as

$$\hat{\mathcal{E}}(\tau) = \sqrt{\hat{\mathcal{E}}_x^2(\tau) + \hat{\mathcal{E}}_y^2(\tau)}. \quad (8)$$

The covariance for a time horizon $\tau$ is computed using $\mathcal{E}_x$ and $\mathcal{E}_y$ for all time steps $t$ and trajectories $i$.

To compare the predicted covariance matrices with the ones obtained from the predicted errors, we use two metrics: (a) the Frobenius norm of the difference

$$\Delta_{\text{F}}(\tau) = \frac{1}{N(T-\tau)} \sum_{i=1}^{N} \sum_{t=1}^{T-\tau} ||P(t,\tau)^{\text{meas}} - P_i(t,\tau)||_{\text{F}}, \quad (9)$$

and (b) the size difference

$$\Delta_{\Lambda}(\tau) = \frac{1}{N(T-\tau)} \sum_{i=1}^{N} \sum_{t=1}^{T-\tau} \Lambda(t,\tau)^{\text{meas}} - \Lambda_i(t,\tau), \quad (10)$$

where $||\cdot||_{\text{F}}$ denotes the Frobenius norm, $P_i(t,\tau)$ is the $\tau$ steps-ahead predicted covariance of scenario $i$ at time $t$, $P(t,\tau)^{\text{meas}}$ is the corresponding measured covariance given the errors, and $\Lambda_i$ and $\Lambda^{\text{meas}}$ are the corresponding products of eigenvalues.

Our LQR-based model propagates all possible bifurcations and the RL-based model propagates trajectories associated with all goals. In order to obtain a sensible error evaluation, when deploying the two approaches we prune all solutions which take different turns than the measured trajectory.

### B. Results

We implemented the algorithms in MATLAB using a sampling time $t_{\text{s}} = 0.1$ s for both methods.

For the RL-based method, since parameter $\alpha$ can be easily tuned, we used values $\alpha \in [20, 50, 100, 1000]$. The discount factor and the cost, instead, need to be tuned together. Since tuning them is not very intuitive and requires one to wait for long computations to run, we did not fine-tune them. The discount factor was set to $\gamma = 0.99$, whereas the classes: road, sidewalk and crosswalk had the rewards $R_{\text{r}} = -3$, $R_{\mathrm{sw}} = -1$ and $R_{\mathrm{cw}} = -1$ respectively. The goals placed in the map had rewards set to $R_{\mathrm{g}} = 0$. The state space was discretized into 20 cm by 20 cm grid cells, and the action space into 24 different directions.

TABLE I
RUNTIME BETWEEN ALGORITHMS AVERAGED OVER 1000 ITERATIONS.

| $\tau$ | 50 | 100 | 150 | 200 |
|---|---|---|---|---|
| LQR | 6.6 ms | 9.3 ms | 14.7 ms | 19.5 ms |
| RL | 1.19 s | 2.14 s | 3.39 s | 4.31 s |

For the LQR-based method we used

$$Q = q \cdot \mathrm{diag}(1\ 1\ 1\ 1),\ R = r \cdot \mathrm{diag}(1\ 1),\ S = 0$$

to tune the controller, and $W = 0.3 \cdot \mathrm{diag}(0.1\ 0.1\ 0.1\ \pi/180)$ as process noise. The ratio between $q$ and $r$ was initially set to 10 and then progressively reduced to 1, 0.1 and, finally 0.02. Also for the LQR we did not fine-tune the cost beyond these tuning steps and the process noise was set once for early simulations and never tuned further. Future research should establish formal methods to identify model parameters systematically through learning or identification methods.

Fig. 8 illustrates the comparison between the LQR-model and the RL-model while plotted against a measured pedestrian trajectory. The predicted motion is similar between the methods, however, the RL-model predicts a motion closer to the edges of the intersection. Our model on the other hand propagates more conservative uncertainty ellipses in comparison to the RL-model for the current model parameters. Fig. 9 shows an example of a non road code abiding trajectory. Because our method is not designed to predict such situations, it predicts that the pedestrian will cross the road closer to the zebra crossing. As time evolves, however, new measurements improve the prediction. Ideally, an intent prediction algorithm should deliver the information that the pedestrian is likely to violate the road code, such that a new edge can be added to our graph for the considered pedestrian.

Fig. 10 shows the average $\ell_2$ error across different prediction horizons in the dataset. As expected, the error increases with longer prediction horizons. It is visible from the figure, that the average error grows less rapidly in the LQR-based approach.

Fig. 11 displays the second error metric, i.e. the Frobenius norm of the difference between the predicted covariance and the measured covariance. Also for this second metric, we notice that the error increases with longer horizons. One can note that the RL-based method has a smaller error for short horizons. However, while the RL-based model slightly underestimates the covariance, the LQR-based approach slightly overestimates it, as shown in Fig. 12. Increasing time-varying process noise could improve the quality of the covariance estimate, and future research should investigate these opportunities.

Finally, the actual runtime of the algorithms is decisive for implementation in autonomous driving applications. Both models were implemented in MATLAB. Neither the RL nor the LQR implementations were optimized. The runtime comparison is therefore only indicative for the intended use in real-time automotive environments. The algorithms were timed by running 1000 simulations with four different prediction horizons. We evaluated the runtime using 100 samples for the RL-based method. From the results presented in Table I, the RL-based method has substantially longer runtimes. Moreover, the LQR timings include the evaluation of the covariance, while the RL timings refer to the evaluation of the samples and do not include the computation of mean, covariance, nor the handling of situations such as the one depicted in Fig. 5.

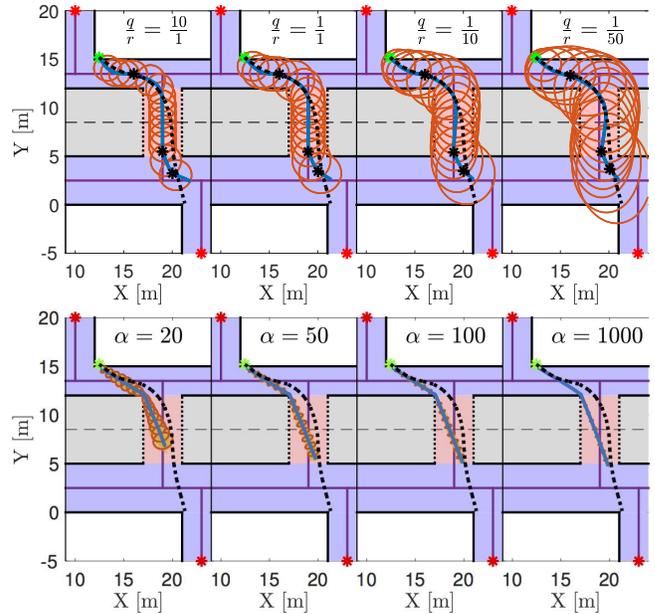

Fig. 8. Predicted motion for the LQR-based model (top plot) and the RL-based model (bottom plot). The color coding matches the one of Fig. 4. The dotted black line represents the measured pedestrian motion, and the black asterisks in the top plot indicate when switching of reference occurs for the LQR-based model.

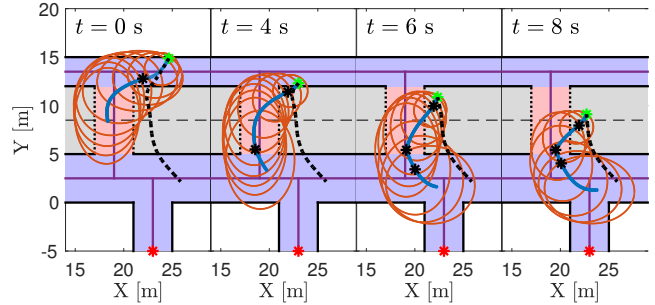

Fig. 9. Predicted motion of the LQR method under different times $t$. The color scheme matches the one of Fig. 8.

Finally, we also compared our model versus a unicycle model which, at every time instant, is linearized around the current predicted average state. This substantial increase in complexity led only to a small improvement in the error metrics.

## V. CONCLUSIONS

We presented a prediction model for pedestrian motions based on simple graph representation of the road geometry.

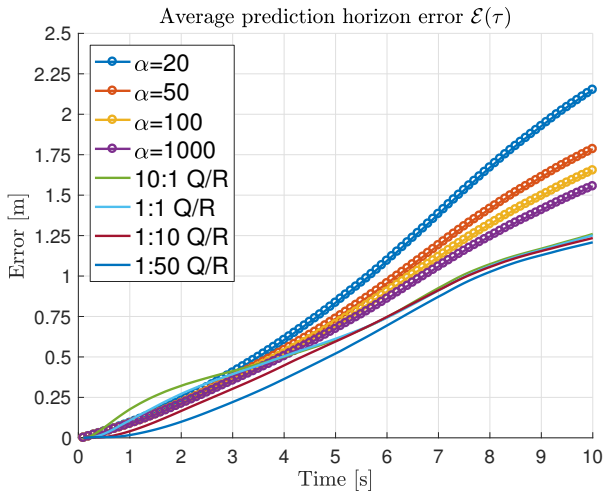

Fig. 10. Performance in terms of $\ell_2$ error.

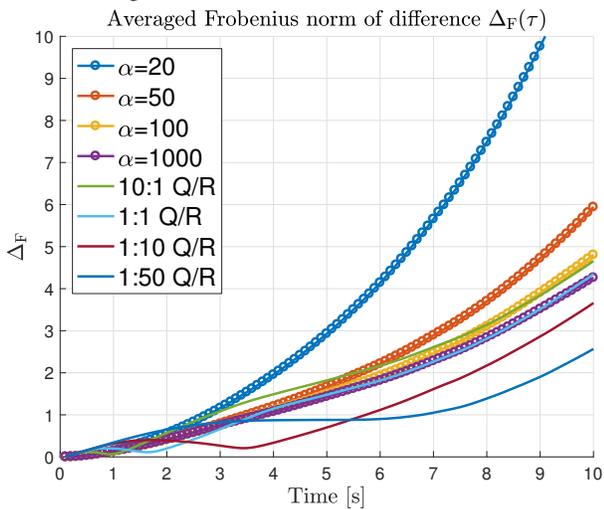

Fig. 11. Covariance error measured by $\Delta_{\mathrm{F}}$.

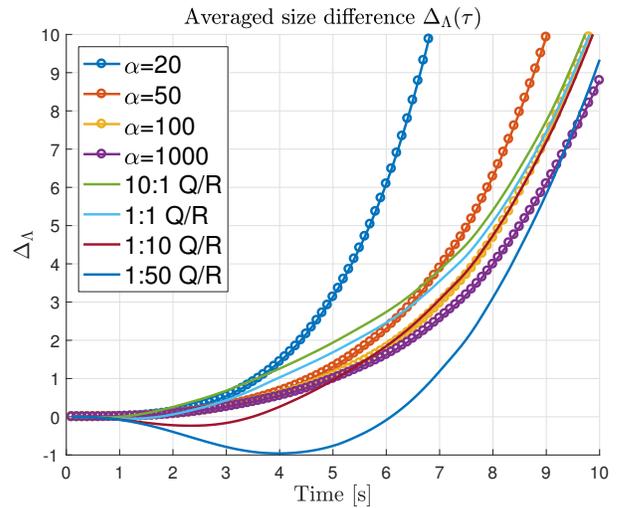

Fig. 12. Covariance error measured by $\Delta_\Lambda$.

Our LQR-based model had an accuracy comparable to a current state-of-the-art RL-model, but with a substantially lower runtime, and a much higher flexibility, since no value function needs to be pre-computed for each road segment. Future research will aim at further improving the promising performance for use in collision avoidance algorithms in automated driving. The results are limited to nominal behavior of single pedestrians not interacting with the environment. The combination of our model with intent prediction algorithms is expected to significantly improve the accuracy of the predictions. Finally, future work will learn or identify the tuning parameters of the model based on real data.